\documentclass[preprint]{daniel_okoh_aastex}
\usepackage{daniel_okoh_spr-astr-addons}
\usepackage{url}

\usepackage{epstopdf}
\RequirePackage{color}

\newcommand{\emaila}

\begin{document}

\title{Formaldehyde and H110$\alpha$ observations towards 6.7 GHz methanol maser sources}
\shorttitle{Okoh et al.: Formaldehyde and H110$\alpha$ Observations towards 6.7 GHz Methanol Maser Sources}
\shortauthors{Okoh et al.}

\author{Daniel. Okoh\altaffilmark{1,2,3}} \and \author{Jarken. Esimbek\altaffilmark{1}}
\and \author{Jian. Jun. Zhou\altaffilmark{1}}
\and \author{Xin. Di. Tang\altaffilmark{1}}
\and \author{Augustine. Chukwude\altaffilmark{3}}
\and \author{Johnson. Urama\altaffilmark{3}}
\and \author{Pius. Okeke\altaffilmark{2}}
\emaila{email: okodan2003@gmail.com}

\altaffiltext{1}{Xinjiang Astronomical Observatory, CAS, Urumqi 830011, China}
\altaffiltext{2}{Center for Basic Space Science, University of Nigeria, Nsukka 410001, Nigeria}
\altaffiltext{3}{Physics \& Astronomy Department, University of Nigeria, Nsukka 410001, Nigeria}

\begin{abstract}
Intriguing work on observations of 4.83 GHz formaldehyde (H$_2$CO) absorptions and 4.87 GHz H110$\alpha$ radio recombination lines (RRLs) towards 6.7 GHz methanol (CH$_3$OH) maser sources is presented. Methanol masers provide ideal sites to probe the earliest stages of massive star formation, while 4.8 GHz formaldehyde absorptions are accurate probes of physical conditions in dense (10$^3$ - 10$^5$ cm$^{-3}$) and low temperature molecular clouds towards massive star forming regions. The work is aimed at studying feature similarities between the formaldehyde absorptions and the methanol masers so as to expand knowledge of events and physical conditions in massive star forming regions. A total of 176 methanol maser sources were observed for formaldehyde absorptions, and formaldehyde absorptions were detected 138 of them. 53 of the formaldehyde absorptions were newly detected. We noted a poor correlation between the methanol and formaldehyde intensities, an indication that the signals (though arise from about the same regions) are enhanced by different mechanisms. Our results show higher detection rates of the formaldehyde lines for sources with stronger methanol signals. The strongest formaldehyde absorptions were associated with IRAS sources and IRDCs that have developed HII regions, and that do not have EGOs.
\end{abstract}



\section{Introduction}
Methanol (CH$_3$OH) masers have been widely established as good tracers for the early stages of high-mass star formation regions \citep{pan11, ell07, cas96}. Hydroxyl and water masers have also been observed to be associated with the early stages of massive star formation regions \citep{bar12}, but they can also be found in the later stages of a star's life \citep{xuy03, ell07b}, making methanol masers unique indicators for the early stages of massive star-formation. Methanol maser emissions have been observed in several transitions \citep{fon10, god07}, the strongest being the 5$_1$ - 6$_0$ A$^+$ line at 6.7 GHz which also happens to be the second strongest Galactic maser of any known molecule \citep{xuy03}.

The 1$_1$$_1$ - 1$_1$$_0$ lines of formaldehyde (H$_2$CO) at 4.8 GHz are mostly seen in absorption towards interstellar dust clouds even when there are no discrete continuum sources behind the clouds \citep{sum75}, so they are good tracers for molecular clouds towards star formation regions \citep{zha12} and accurate probes of physical conditions in dense (10$^3$ - 10$^5$ cm$^{-3}$) molecular clouds \citep{man93}. Several surveys \citep{dow80, duz11, han11, wug11} reveal that HII regions are associated with 4.8 GHz H$_2$CO absorptions, and since HII regions are known to be regions where star formations are taking place, the surveys suggest some association of 4.8 GHz H$_2$CO absorptions with star forming regions.

H$_2$CO and CH$_3$OH are also believed to be formed in the interstellar medium by successive hydrogenation of carbon monoxide (CO) on ice surfaces \citep{cha97, wat02, woo02} through the sequence of reactions: CO $\rightarrow$ HCO $\rightarrow$ H$_2$CO $\rightarrow$ CH$_3$O $\rightarrow$ CH$_3$OH. \cite{wat02} showed through their atomic hydrogen beam experiment that both molecules were efficiently produced by the hydrogenation of CO in H$_2$O-CO ice at 10 K. Values for percentage yields from their experiment compared favorably with observed abundances in objects toward high-mass protostars, and further confirms that the addition of H-atom to CO in molecular clouds is most likely to account for the formation of H$_2$CO and CH$_3$OH in those regions. In a related survey of deuterated formaldehyde and methanol towards low-mass class 0 protostars, \cite{par06} detected HDCO, D$_2$CO and CH$_2$DOH in all 6 sources surveyed, and though they detected CH$_3$OD and CHD$_2$OH respectively in only 2 and 3 of the 6 sources, their results reveal an extent of co-existence between the formaldehyde and methanol molecules. Recent work by \cite{oko13} using the k-means algorithm to cluster H$_2$CO and CH$_3$OH observations also reveal a strong sense of co-existence between the two molecules.

Following the detection of formaldehyde absorptions at 4830 MHz by \cite{sny69}, several observations has been conducted at different times and stations with varying aims. \cite{whi70} observed 34 galactic sources using the Parkes 64-m reflector with the main aim of investigating the galactic distribution of the absorption, and to resolve kinematic distance ambiguities (KDAs) resulting from an earlier survey of H109$\alpha$ RRLs by \cite{wil70}. \cite{gar73} also used the Parkes telescope to make 6 cm formaldehyde absorption observations towards the Carina Nebula with the aim of investigating how the results contributed to the understanding of activities in the nebula, and they concluded that the results do not show significant contribution to the understanding of the nebula. A similar survey by \cite{dic77} revealed that there was no detailed correlation between their observed H$_2$CO absorptions and the dust in nebula IC 1318 b-c. \cite{die73} surveyed 381 dust clouds from the Lynds's Catalogue of Dark Nebulae for formaldehyde absorptions, and they noted that the more opaque the dust clouds were, the lower their central excitation temperatures, and proposed a model of typical formaldehyde cloud associated with dust clouds which supports the collisional mechanism for cooling the molecules. Also using the Nanshan telescope, \cite{tan13} illustrated that there was a good morphological relation between $^{13}$CO and the H$_2$CO absorptions on a large scale, and that the two tracers systematically provide consistent views  of the dense regions.

In this work, we seek to quantitatively study relations between the 6.7 GHz methanol masers and the 4.8 GHz formaldehyde absorptions in order to present more empirical results on activities and conditions in massive star formation regions.

\section{Sample and Observations}
\subsection{Sample}
We first made a catalogue of 6.7 GHz methanol maser sources from the works of \cite{xuy03}, \cite{pes05}, \cite{xuy09}, and \cite{gre10}. A total of 1,888 sources were listed, and considering that the declination limit for reliable observation from our station and telescope is -28$^{\circ}$, we removed sources with declination angles less than -28$^{\circ}$ to be left with 1,066 sources. Next we removed sources that were duplicated in two or more of the four papers so as to retain a listing of only unique sources, this brought the number down to 766. And finally, considering that the spatial resolution of our 25-m telescope is ~10 arc minutes at the 6-cm observation wavelength, we removed sources that were within 10 arc minutes of each other, retaining the ones with the greatest methanol maser intensities, and only those with intensities greater than 5 Jy, bringing the final number of sources to a total of 176.

\subsection{Observations}
Observations were made using the Nanshan 25-m radio telescope located at 87$^{\circ}$E, 43$^{\circ}$N, altitude 2080 m, and operated by the Xinjiang Astronomical Observatory, CAS, Urumqi, China. The telescope's C-band cryogenic receiver (23K system temperature) was used together with the newly installed Digital Filter Bank (DFB) system. The observation frequencies were centered around 4851.9102 MHz with a bandwidth of 64 MHz in 8192 channels corresponding to spectral resolution of 7.8 kHz and velocity resolution of 0.48 km s$^{-1}$. Formaldehyde absorption lines and H110$\alpha$ recombination lines were simultaneously observed at frequencies of 4829.6594 MHz and 4874.1570 MHz respectively. The system's spatial resolution at this frequency band is about 10 arc minutes, and the Degrees-per-flux unit (DPFU) value is 0.116 K Jy$^{-1}$. A diode noise source was used to calibrate the spectrum and the flux error was about 10\%. The telescope's pointing and tracking accuracy was better than 15 arc seconds, and the beam efficiency was 65\%. The ON/OFF mode of observation was used, and to optimize observation time, we first made 24 minutes integration time observations of each source and an additional 24 minutes integration time for sources with signal-to-noise (S/N) ratio less than 3. Observations were made between May and December 2012, and W3 was always used to verify that the system works well.

\section{Results}
\subsection{Data Reduction and Analysis}
Data reduction, visualization, and spectral analysis were done using the Continuum and Line Analysis Single-dish Software (CLASS) which is part of the Grenoble Image and Line Data Analysis Software (GILDAS) developed by IRAM Grenoble. Weighted spectra were averaged and calibrated, the baselines were removed, and Gaussian profiles were fitted to the spectra. For sources where we found formaldehyde absorptions and H110$\alpha$ RRLs, we derived the line parameters from the Gaussian fits.

\subsection{Results}
Table 1 is a listing of all 176 methanol sources observed. Methanol velocities and intensities are shown in columns 4 and 5 as obtained from references cited in column 6. Columns 7 and 8 are respectively flags that indicate whether or not we obtained H$_2$CO and H110$\alpha$ signals from observing the sources; flag 0 means signals were not detected, while 1 means signals were detected. Using a catalogue of observed H$_2$CO absorptions constructed from literature sources illustrated in Table 2, we assign a flag of 2 to H$_2$CO absorption sources that have not been previously observed. The results show that 53 of the H$_2$CO absorptions are newly detected in this work.

Tables 3 and 4 respectively contain the line parameters of the H$_2$CO and H110$\alpha$ observations. Plots of CH$_3$OH versus H$_2$CO line intensities are shown on a log-log scale in figure 1(a), while figure 1(b) is a histogram illustration of velocity differences between the formaldehyde and methanol sources. We noted a weak correlation (-0.03 correlation coefficient) between the line intensities.

\section{Discussions}
\subsection{H$_2$CO and CH$_3$OH}
The results of figure 1 show a weak correlation between the formaldehyde and methanol line intensities, an indication that the mechanisms that enhance their intensities are likely different \citep{oko12}. 6.7 GHz methanol masers (5$_1$ - 6$_0$ A$^+$ transitions) are known to be directly associated with the birth place of massive stars and to be pumped radiatively by infrared radiation from nearby warm dust \citep{pes05} while 4.8 GHz formaldehyde lines (1$_1$$_1$ - 1$_1$$_0$ transitions) are usually observed in absorption against the cosmic background, and are believed to be collisionally pumped \citep{tow69}. The beam difference between the formaldehyde and methanol observations is a limitation to lucid conclusions in this work.

To present more studies on sources with larger differences between the methanol and formaldehyde velocities, we first compute the standard deviation of differences between all methanol and formaldehyde velocities, and then note sources with velocity differences higher than the standard deviation. These are sources with velocity differences outside the $\pm$ 16 kms$^-$$^1$ region of figure 1(b). For those sources, we compared their methanol and formaldehyde velocities with those of observed CO sources and noted the following: (i) Some of them had corresponding CO velocities for the different methanol and formaldehyde velocities. For example, the source G18.66+0.03 has methanol velocity at 80.2 kms$^-$$^1$ and formaldehyde velocity at 45.7 kms$^-$$^1$, we noted observed CO velocities at 80.2 kms$^-$$^1$ and at 45.5 kms$^-$$^1$ indicating two molecular regions along those lines of sight. (ii) Others had CO velocities relatively closer to the formaldehyde velocities than the methanol velocities. For example G10.32-0.26 has methanol velocity at 39.0 kms$^-$$^1$ and formaldehyde velocity at 10.6 kms$^-$$^1$, there is an observed CO velocity at 10.0 kms$^-$$^1$, indicating that methanol was observed at a clearly different velocity from formaldehyde and CO. These are probabely unrelated observations of the high-density core tracer (methanol) and  low-density core tracers (formaldehyde and CO). We have ignored these sources in the intensity studies. CO data were obtained from the Millimeter Wave Radio Astronomy Database (http://www.radioast.csdb.cn/indexenglish.php).

Despite the poor correlation we noted between intensities of the two lines, we observed that the H$_2$CO detection rates were generally higher towards stronger methanol sources. We put the 176 methanol sources into 4 categories: (a) I $\geq$ 100 Jy, (b) 50 Jy $\leq$ I $<$ 100 Jy, (c) 10 Jy $\leq$ I $<$ 50 Jy, and (d) 5 Jy $\leq$ I $<$ 10 Jy. Table 5 illustrates the detection rates for each category. The table shows that the detection rates were generally higher towards methanol sources with greater signal intensities. For all categories, the detection rates were higher than 60\%, and in general, the overall detection rate was about 80\%. The result suggests very close associations between the methanol and formaldehyde lines, and further supports the proposals of \cite{cha97} and \cite{wat02} that both H$_2$CO and CH$_3$OH molecules are formed in interstellar medium by the same process of successive hydrogenation of CO. This is also in agreement with the works of \cite{tan13} who observed good morphological relation between  H$_2$CO and the $^{13}$CO line.

\subsection{H110$\alpha$}
Of the 176 methanol maser sources observed, we detected H110$\alpha$ RRLs in 56 of them. In all the sources where we detected H110$\alpha$ RRLs, we also detected H$_2$CO absorption lines except for G79.75+0.99, G23.80+0.40, and G18.34+1.77. Similar work by others \citep{han11, duz11} also reported a few cases of H110$\alpha$ RRLs detections that were not associated with H$_2$CO absorption lines. That we detected H$_2$CO absorption lines in nearly all H110$\alpha$ RRLs show a significant level of association of H$_2$CO with  HII regions as also demonstrated in the works of \cite{dow80}. A plot of the observed H110$\alpha$ versus H$_2$CO intensities reveals a seemingly linear relationship between the two; this is shown in figure 2, and a least square linear fit to the data gives
I(H110$\alpha$)=0.405I(H$_2$CO)+0.066 (Jy), where I(H110$\alpha$) is H110$\alpha$ intensity, and I(H$_2$CO) is absolute H$_2$CO intensity.

A further look at the CH$_3$OH versus H$_2$CO intensity plot for sources associated with and without H110$\alpha$ signals (figure 3(a)) reveals that the majority of high intensity H$_2$CO sources (with absolute intensities greater than 3 Jy) are associated with HII regions, an observation that could be explained to be due to the collisional pumping model proposed by \cite{tow69}; H$_2$CO absorptions are strongest at HII regions owing to collisional pumping by the ionized gas in those high temperature regions \citep{zha12}. We obtained a mean intensity value of 2.42 Jy for H$_2$CO absorptions associated with H110$\alpha$ RRLs, and a mean intensity value of 0.91 Jy for those not associated with H110$\alpha$ RRLs. H$_2$CO absorptions associated with H110$\alpha$ RRLs had intensities in the range of 0.25 Jy - 22.93 Jy while those not associated with H110$\alpha$ RRLs had intensities in the range of 0.16 Jy - 3.34 Jy, an indication that the H$_2$CO absorptions are conspicuously affected by HII regions as earlier reported by \cite{dow80}, and as figure 2 depicts.

\subsection{EGOs}
Extended Green Objects (EGOs) are a recent unveiling of probable candidates for massive young stellar objects (MYSOs). We used the \cite{cyg08} catalogue of 302 EGOs to identify, in our observations, H$_2$CO absorptions that were associated with EGOs, and we found 28 of them. A plot of CH$_3$OH versus H$_2$CO intensities for sources associated with and without the EGOs (figure 3 (b)) revealed that the EGOs were mostly associated with the relatively weak CH$_3$OH masers and H$_2$CO absorptions; the EGO-associated sources were noted to be mostly confined within the region of the graph with CH$_3$OH  maser intensities $<$ 400 Jy and H$_2$CO absolute intensities $<$ 3 Jy. EGOs are well established to be the very earliest stages of massive stellar objects \citep{hej12, cyg08}, with their remarkable 4.5 $\mu$m emissions being primarily due to collisionally excited lines of molecular hydrogen \citep{deb10}. \cite{cha09} proposed an evolutionary sequence in which clumps first form EGOs before developing HII regions, revealing that the CH$_3$OH  masers and H$_2$CO absorptions that are associated with EGOs most likely emanate from regions that are yet to develop the bright infrared radiations and HII regions required to respectively enhance their excitations.

\subsection{IRDCs}
Infrared dark clouds (IRDCs) are objects seen in absorption against the bright diffuse Galactic mid-infrared background and their cold dense nature with substantial masses strongly suggests IRDCs are sites for massive star and/or cluster formation \citep{liu11, kim10}. Using the Spitzer catalogue of IRDCs \citep{per09}, we found that 93 of the 138 sources where we noted H$_2$CO absorptions were associated with IRDCs. A plot of CH$_3$OH versus H$_2$CO intensities for sources associated with and without IRDCs is shown in figure 3 (c). The figure suggests that the very strong H$_2$CO absorptions are associated with IRDCs, agreeing with the findings of \cite{bat10} that several IRDCs had embedded UC HII regions and collaborating our earlier remarks that H$_2$CO absorption intensities were significantly enhanced by HII regions. According to the \cite{cha09} evolutionary sequence, clumps found in IRDCs evolve from "quiescent" (with no IR Spitzer emission) to "intermediate" (with either an EGO or a 24 $\mu$m source, but not both) to "active" (with both an EGO and a 24 $\mu$m source) to "red" (with an HII region). By comparing figures 3(b) and 2(b), we confirm that the very strong H$_2$CO absorptions are associated with the "red" IRDCs, which have embedded HII regions. A similar plot for observations that are associated with and without IRAS sources (figure 3 (d)) also confirms that the strong H$_2$CO absorptions are associated with IRAS sources that are in HII regions. We obtained the IRAS catalogue from the Harvard IRAS Source Catalogue (http://tdc-www.harvard.edu/software/catalogs/iras.html).

Table 6 is a summary that indicates which of the individual formaldehyde sources are associated with  H110$\alpha$, EGO, IRDC and IRAS sources. The number 1 denotes an association with the source while the number 0 denotes a non-association.

\section{Conclusions}
Both 6.7 GHz CH$_3$OH masers and 4.8 GHz H$_2$CO absorptions are established to be associated with massive star formation regions. We have simultaneously made observations of 4.83 GHz H$_2$CO absorptions and 4.87 GHz H110$\alpha$ RRLs towards 176 6.7--GHz CH$_3$OH maser sources, and detected 4.8 GHz H$_2$CO absorptions in about 80\% of them, 53 of which were newly detected.

The detection rate of H$_2$CO absorptions towards CH$_3$OH maser sources was generally high, supporting earlier expectations of associations between the two molecular lines, and that both molecules are formed in interstellar medium via the same mechanism. Detection rates were generally higher towards the stronger CH$_3$OH maser sources, but there was no good correlation between the formaldehyde and methanol line intensities, an indication that (within the limitations of errors introduced by the beam differences) the mechanisms that enhance their intensities are different.

A good correlation was noted between the H110$\alpha$ RRL and H$_2$CO absorption line intensities, indicating that the continuum from HII regions enhance the H$_2$CO intensities. We detected H110$\alpha$ RRLs towards less than half of the H$_2$CO absorption sources, but the strongest H$_2$CO absorptions were found to be associated with H110$\alpha$ RRLs, supporting the idea that H$_2$CO absorptions in HII regions are enhanced by collisional excitations in those regions.

The strongest H$_2$CO absorptions were found to be associated with IRDCs and IRAS sources, but not with EGOs, supporting earlier propositions of an evolutionary sequence in which EGOs form in IRDCs before strong IR radiations and HII regions are developed. We conclude that these strong H$_2$CO absorptions emanate from the "red" IRDCs, which have developed embedded HII regions.

\acknowledgments
We are immensely grateful to TWAS (the Academy of Science for the Developing World) and CAS (the Chinese Academy of Science) for their generous travel and financial supports during the periods of this research work. The work was also funded by the National Natural Science foundation of China under grant 10778703, and partly supported by the China Ministry of Science and Technology under State Key Development Program for Basic Research (2012CB821800) and The National Natural Science foundation of China under grant 11373062, 11303081 and 10873025.


\begin{thebibliography}{}

\bibitem[Araya et al.(2002)]{ara02} Araya, E., Hofner, P., Churchwell, E., Kurtz, S.: \apjs, 138, 63 (2002)
\bibitem[Araya et al.(2004)]{ara02} Araya, E., Hofner, P., Linz, H., Sewilo, M.,Watson, C., Churchwell, E., Olmi, L., Kurtz, S.: \apjs, 154, 579 (2004)
\bibitem[Araya et al.(2007)]{ara02} Araya, E., Hofner, P., Goss, W. M., Linz, H., Kurtz, S., Olmi, L.: \apjs, 170, 152 (2007)
\bibitem[Bartkiewicz \& Langevelde(2012)]{bar12} Bartkiewicz, A., Langevelde, H. J.: IAUS 287 (2012)
\bibitem[Battersby et al.(2010)]{bat10} Battersby, C., Bally, J., Jackson, J. M., Ginsburg, A., Shirley, Y., Schlingman, W., Glenn, J.: \apj, 721, 222 (2010)
\bibitem[Boyce et al.(1989)]{boy89} Boyce, P. J., Cohen, R. J., Dent, W. R. F.: \mnras, 239, 1013 (1989)
\bibitem[Caswell(1996)]{cas96} Caswell, J. L.:  \mnras, 279, 79 (1996)
\bibitem[Caswell \& Haynes(1987)]{cas87} Caswell, J. L., Haynes, R. F.: \aap, 171, 261 (1987)
\bibitem[Chambers et al.(2009)]{cha09} Chambers, E. T., Jackson, J. M., Rathborne, J. M., Simon, R.: \apjs, 181, 360 (2009)
\bibitem[Charnley et al.(1997)]{cha97} Charnley, S. B., Tielens, A. G. G. M., Rodgers, S. D.: \apj, 482, L203 (1997)
\bibitem[Cyganowski et al.(2008)]{cyg08} Cyganowski, C. J, Whitney, B. A, Holden, E., Braden, E., Brogan, C. L., Churchwell, E., Indebetouw, R., Watson, D. F., Babler, B. L., Benjamin, R., Gomez, M., Meade, M. R., Povich, M. S, Robitaille, T. P., Watson, C.: \aj, 136, 2391 (2008)
\bibitem[De Buizer \& Vacca(2010)]{deb10} De Buizer, J. M., Vacca, W. D.: \aj, 140, 196 (2010)
\bibitem[Dickel et al.(1977)]{dic77} Dickel, H. R., Seacord, A. W., Gottesman, S. T.: \apj, 218, 133 (1977)
\bibitem[Dieter(1973)]{die73} Dieter, N. H.: \apj, 183, 449 (1973)
\bibitem[Downes et al.(1980)]{dow80} Downes, D., Wilson, T.L., Bieging, J., Wink, J.:  \aaps, 40, 379 (1980)
\bibitem[Du et al.(2011)]{duz11} Du, Z. M., Zhou, J. J., Esimbek, J., Han, X. H., Zhang, C. P.:  \aap, 532, 127 (2011)
\bibitem[Duncan et al.(1987)]{dun87} Duncan, R. A., Forster, J. R., Gardner, F. F., Whiteoak, J. B.: \mnras, 224, 721 (1987)
\bibitem[Ellingsen(2007)]{ell07} Ellingsen, S. P.: \mnras, 377, 571E (2007)
\bibitem[Ellingsen et al.(2007)]{ell07b} Ellingsen, S. P., Voronkov, M. A., Cragg, D. M., Sobolev, A. M., Breen, S. L., Godfrey, P. D.: Proc. IAU, 242, 213 (2007)
\bibitem[Fontani et al.(2010)]{fon10} Fontani, F., Cesaroni, R., Furuya, R. S.:  \aap, 517, A56 (2010)
\bibitem[Gardner et al.(1973)]{gar73} Gardner, F. F., Dickel, H. R., Whiteoak, J. B.:  \aap, 23, 51 (1973)
\bibitem[Gardner \& Whiteoak(1984)]{gar84} Gardner, F. F., Whiteoak, J. B.: \mnras, 210, 23 (1984)
\bibitem[Ginsburg et al.(2011)]{gin11} Ginsburg, A., Darling, J., Battersby, C., Zeiger, B., Bally, J.: \apj, 736, 149 (2011)
\bibitem[Goddi et al.(2007)]{god07} Goddi, C., Moscadelli, L., Sanna, A., Cesaroni, R., Minier, V.:  \aap, 461, 1027 (2007)
\bibitem[Goss et al.(1984)]{gos84} Goss, W. M., Kalberla, P. M. W., Dickel, H. R.: \aap, 139, 317 (1984)
\bibitem[Goss et al.(1980)]{gos80} Goss, W. M., Manchester, R. N., Brooks, J. W., Sinclair, M. W., Manefield, G. A.: \mnras, 191, 533 (1980)
\bibitem[Green et al.(2010)]{gre10} Green, J. A., Caswell, J. L., Fuller, G. A., Avison, A., Breen, S. L., Elligsen, S. P., Gray, M. D., Pestalozzi, M., Quinn, L., Thompson, M. A., Voronkov, M. A.: \mnras, 409, 913 (2010)
\bibitem[Han et al.(2011)]{han11} Han, X. H., Zhou, J. J., Esimbek, J., Wu, G., Gao, M. F.: RAA, 11, 156 (2011)
\bibitem[He et al.(2012)]{hej12} He, J. H., Takahashi, S., Chen, X.: \apjs, 202, 1 (2012)
\bibitem[Heithausen et al.(1987)]{hei87} Heithausen, A., Mebold, U., de Vries, H.W.: \aap, 179, 263 (1987)
\bibitem[Hoffman et al.(2007)]{hof07} Hoffman, I. M., Goss, W. M., Palmer, P.: \apj, 654, 971 (2007)
\bibitem[Javanaud(1979)]{jav79} Javanaud, C.: \mnras, 188, 203 (1979)
\bibitem[Kim et al.(2010)]{kim10} Kim, G., Lee, C. W., Kim, J., Lee, Y., Ballesteros-Paredes, J., Myers, P., Kurtz, S.: JKAS, 43, 9 (2010)
\bibitem[Kogul et al.(1989)]{kog89} Kogul, A., Smoot, G. F., Bennett, C. L., Petuchowski, S. J.: \apj, 346, 763 (1989)
\bibitem[Kutner \& Thaddeus(1971)] {kut71} Kutner, M., Thaddeus, P.: \apj, 168, L67 (1971)
\bibitem[Liu et al.(2011)]{liu11} Liu, S. Y., Su, Y. N., NRO 45M Survey Team: IAUS 280, 55, 2 (2011)
\bibitem[Loren(1981)]{lor81} Loren, R. B.: \aj, 86, 1 (1981)
\bibitem[Lucas et al.(1976)]{luc76} Lucas, R., Encrenaz, P. J., Falgarone, E. G.: \aap, 51, 469 (1976)
\bibitem[Mangum \& Wootten(1993)]{man93} Mangum, J. G., Wootten, A.: \apjs, 89, 123 (1993)
\bibitem[Martin-Pitando et al.(1985)]{mar85} Martin-Pintado, J., Wilson, T. L., Johnston, K. J., Henkel, C.: \apj, 299, 386 (1985)
\bibitem[Mehringer et al.(1995)]{meh95} Mehringer, D. M., Palmer, P., Goss, W. M.: \apjs, 97, 497 (1995)
\bibitem[Minn \& Greenberg(1973)]{min73} Minn, Y. K., Greenberg, J. M.: \aap, 22, 13 (1973)
\bibitem[Minn \& Greenberg(1979)]{min79} Minn, Y. K., Greenberg, J. M.: \aap, 77, 37 (1979)
\bibitem[Okoh et al.(2012)]{oko12} Okoh, D., Esimbek, J., Zhou, J. J., Tang, X. Chukwude, A., Urama, J., Okeke, P.: Proc. IAU, 8, 111 (2012)
\bibitem[Okoh et al.(2013)]{oko13} Okoh, D., Esimbek, J., Zhou, J. J., Tang, X. Chukwude, A., Urama, J., Okeke, P.: J. Space Sci. Tech., 2, 1 (2013)
\bibitem[Pandian et al.(2011)]{pan11} Pandian, J. D., Momjian, E., Xu, Y., Menten, K. M., Goldsmith, P. F.: \apj, 730, 55 (2011)
\bibitem[Parise et al.(2006)]{par06} Parise, B., Ceccarelli, C., Tielens, A. G. G. M., Castets, A., Caux, E., Lefloch, B., Maret, S.:  \aap, 453, 949 (2006)
\bibitem[Peretto \& Fuller(2009)]{per09} Peretto, N., Fuller, G. A.:  \aap, 505, 405 (2009)
\bibitem[Pestalozzi et al.(2005)]{pes05} Pestalozzi, M., Minier, V., Booth, R.:  \aap, 432, 737 (2005)
\bibitem[Pipenbrink \& Wendker(1988)]{pip88} Pipenbrink, A., Wendker, H. J.: \aap, 191, 313 (1988)
\bibitem[Poeppel et al.(1983)]{poe83} Poeppel, W. G. L., Rohlfs, K., Celnik, W.: \aap, 126, 152 (1983)
\bibitem[Reynoso \& Goss(2002)]{rey02} Reynoso, E. M., Goss, W. M.: \apj, 575, 871 (2002)
\bibitem[Rickard et al.(1977)]{ric77} Rickard, L. J., Palmer, P., Buhl, D., Zuckerman, B.: \apj, 213, 654 (1977)
\bibitem[Rieu \& Pankonin(1977)]{rie77} Rieu, N. Q., Pankonin, V.: \aap, 60, 313 (1977)
\bibitem[Rodriguez et al.(2006)]{rod06} Rodriguez, M. I., Allen, R. J., Loinard, L., Wiklind, T.: \apj, 652(2), 1230 (2006)
\bibitem[Sandqvist \& Lindroos(1976)]{san76} Sandqvist, A., Lindroos, K. P.: \aap, 53, 179 (1976)
\bibitem[Sandqvist et al.(1988)]{san88} Sandqvist, A., Tomboulides, H., Lindblad, P. O.: \aap, 205, 225 (1988)
\bibitem[Scoville \& Solomon(1972)]{sco72} Scoville, N. Z., Solomon, P. M.: \apj, 172, 335 (1972)
\bibitem[Snyder et al.(1969)]{sny69} Snyder, L. E., Buhl, D., Zuckerman, B., Palmer, P.: \prl, 22, 679 (1969)
\bibitem[Sume et al.(1975)]{sum75} Sume, A., Downes, D., Wilson, T. L.:  \aap, 39, 435 (1975)
\bibitem[Tang et al.(2013)]{tan13} Tang, X. D., Esimbek, J., Zhou, J. J., Wu, G., Ji, W. G., Okoh, D.:  \aap, 551, 28 (2013)
\bibitem[Townes \& Cheung(1969)]{tow69} Townes, C. H., Cheung, A. C.: \apj, 157, L103 (1969)
\bibitem[Vanden-Bout et al.(1983)]{van83} Vanden-Bout, P. A., Snell, R. L., Wilson, T. L.: \aap, 118, 337 (1983)
\bibitem[Wadiak et al.(1988)]{wad88} Wadiak, E. J., Rood, R. T., Wilson, T. L.: \apj, 324, 931 (1988)
\bibitem[Watanabe \& Kouchi(2002)]{wat02} Watanabe, N., Kouchi, A.: \apj, 571, L173 (2002)
\bibitem[Wendker et al.(1983)]{wen83} Wendker, H. J., Schramm, K. J., Dieckvoss, C.: \aap, 121, 69 (1983)
\bibitem[Whiteoak \& Gardner(1970)]{whi70} Whiteoak, J. B., Gardner, F. F.: \apjl, 5, 5 (1970)
\bibitem[Whiteoak \& Gardner(1974)]{whi74} Whiteoak, J. B., Gardner, F. F.: \aap, 37, 389 (1974)
\bibitem[Wilson(1972)]{wil72} Wilson, T. L.: \aap, 19, 354 (1972)
\bibitem[Wilson et al.(1970)]{wil70} Wilson, T. L., Mezger, P. G., Gardner, F. F, Milne, D. K.:  \aap, 6, 364 (1970)
\bibitem[Woon(2002)]{woo02} Woon, D. E.: \apj, 569, 541 (2002)
\bibitem[Wu et al.(2011)]{wug11} Wu, G., Esimbek, J., Zhou, J., Han, X.: RAA, 11, 63 (2011)
\bibitem[Xu et al.(2009)]{xuy09} Xu, Y., Voronkov, M. A., Pandian, J. D., Li, J. J., Sobolev, A. M., Brunthaler, A., Ritter, B., Menten, K. M.:  \aap, 507, 1117 (2009)
\bibitem[Xu et al.(2003)]{xuy03} Xu, Y., Zheng, X., Jiang, D.: \cjaa, 3, 49 (2003)
\bibitem[Young et al.(2004)]{you04} Young, K. E., Lee, J., Evans, N. J., Goldsmith, P. F., Doty, S. D.: \apj, 614, 252 (2004)
\bibitem[Zylka et al.(1992)]{zyl92} Zylka, R., Guesten, R., Henkel, C., Batrla, W.: \aaps, 96, 525 (1992)
\bibitem[Zhang et al.(2012)]{zha12} Zhang, C. P., Esimbek, J., Zhou, J. J., Wu, G., Du, Z. M.: \apss, 337, 283 (2012)

\end{thebibliography}

\onecolumn

\begin{figure*}[t]
\centering
\includegraphics[width=17cm]{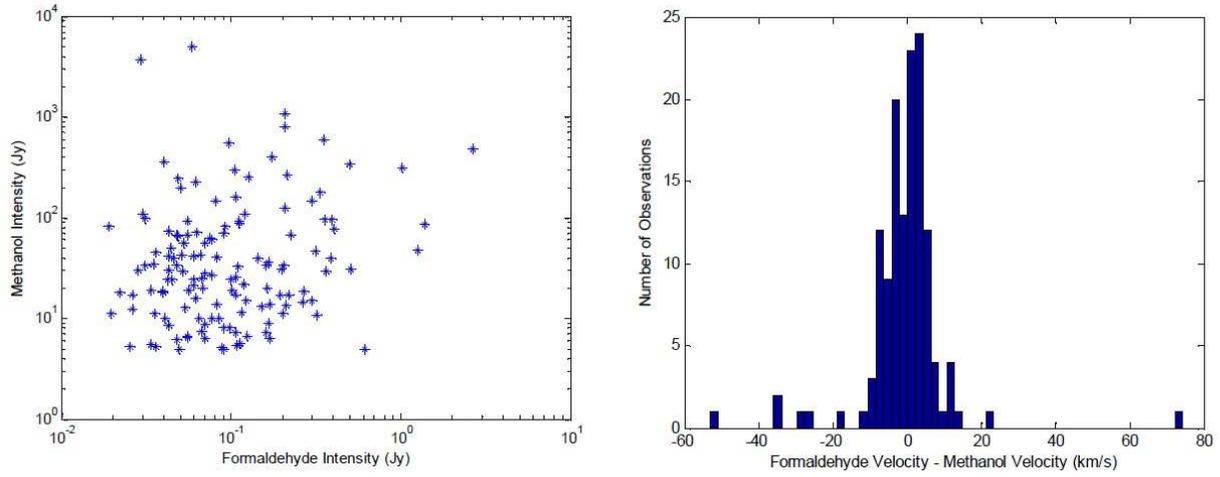}
\caption{(a) Log-log Line intensity plots of formaldehyde absorptions versus methanol masers [Formaldehyde absorption intensities are represented in absolute values], and (b) Histogram representation of velocity differences between the formaldehyde and methanol sources.}
\end{figure*}

\begin{figure}[t]
\centering
\includegraphics[width=10cm]{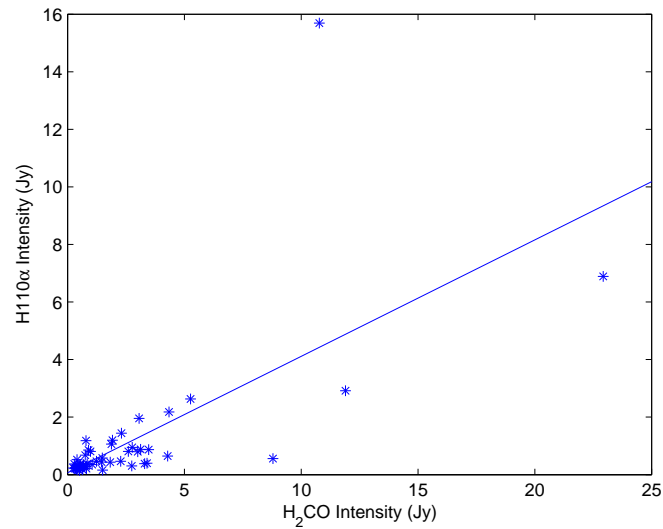}
\caption{Observed H110$\alpha$ versus absolute H$_2$CO intensities.}
\end{figure}

\begin{figure*}[t]
\centering
\includegraphics[width=17cm]{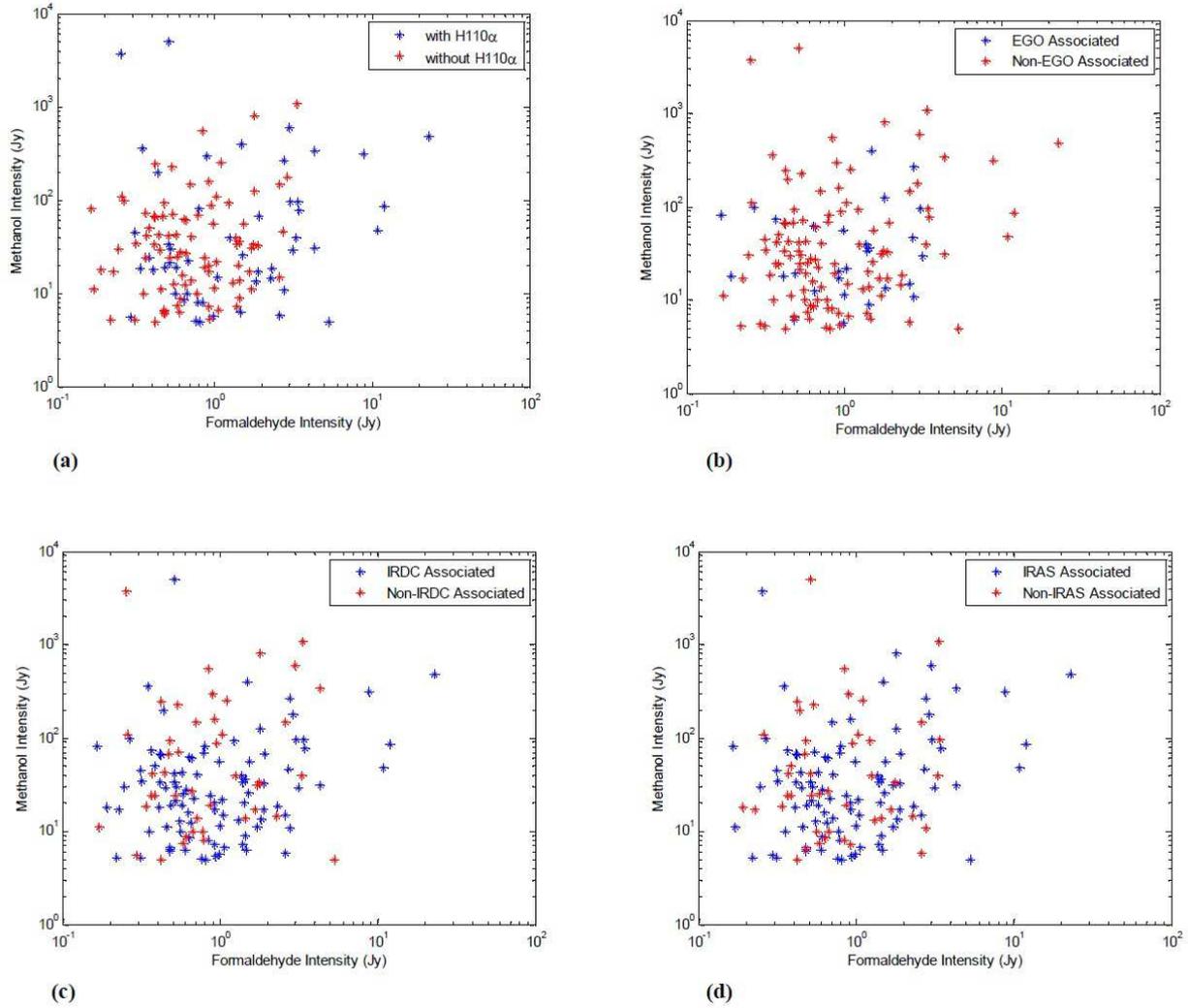}
\caption{Log-log plots of CH$_3$OH versus H$_2$CO absolute intensities for sources associated with and without (a) H110$\alpha$ RRLs, (b) EGOs, (c) IRDCs, and (d) IRAS sources.}
\end{figure*}

\begin{figure}[h]
\centering
\includegraphics[width=17cm, height=20cm]{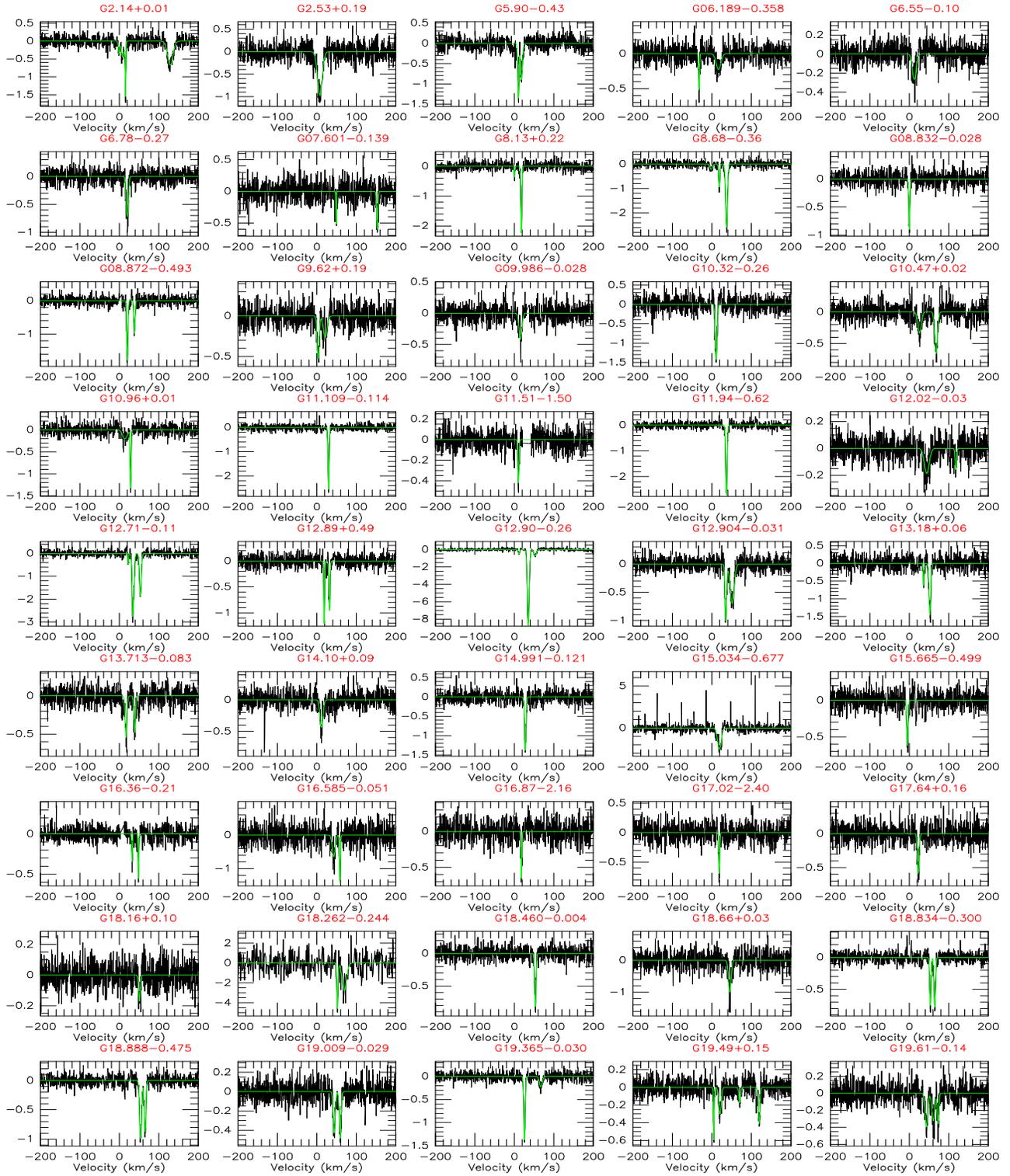}
\caption{Observed Formaldehyde Absorption Spectral Lines (The vertical axes are Line Intensities (in Jy)).}
\end{figure}

\addtocounter{figure}{-1}
\begin{figure}[h]
\centering
\includegraphics[width=17cm, height=20cm]{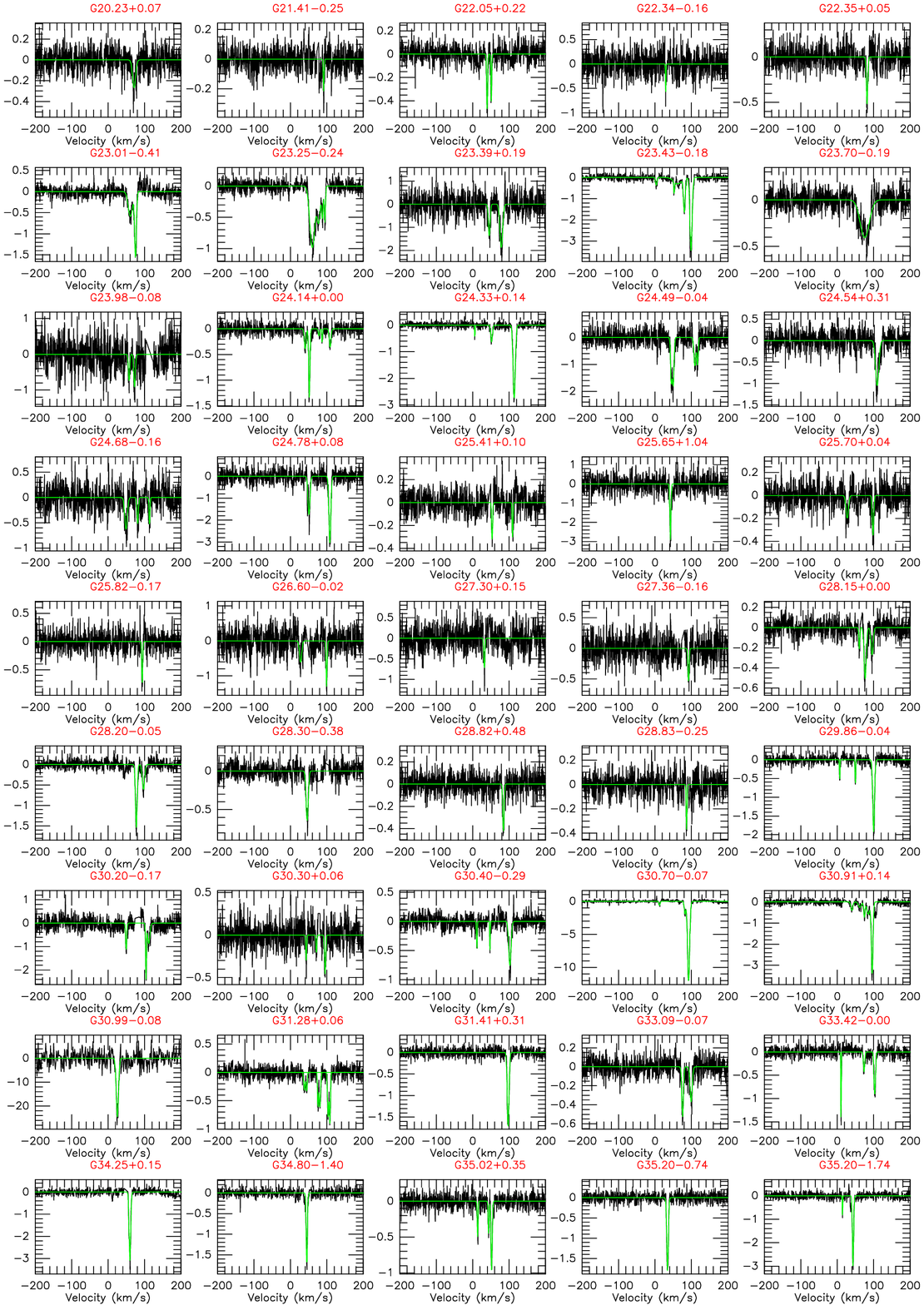}
\caption{--Continued.}
\end{figure}

\addtocounter{figure}{-1}
\begin{figure}[h]
\centering
\includegraphics[width=17cm, height=20cm]{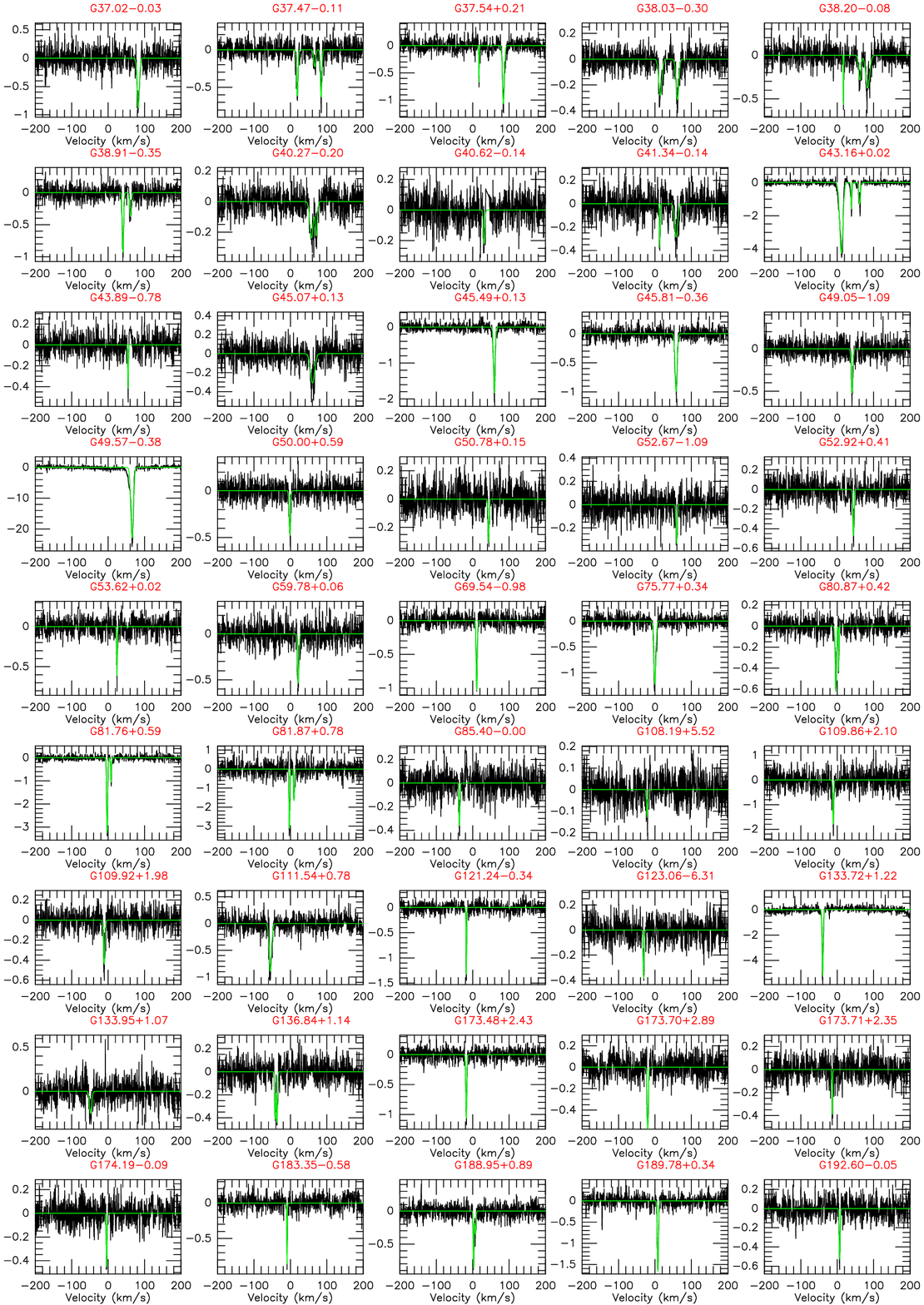}
\includegraphics[width=10cm, height=2.3cm]{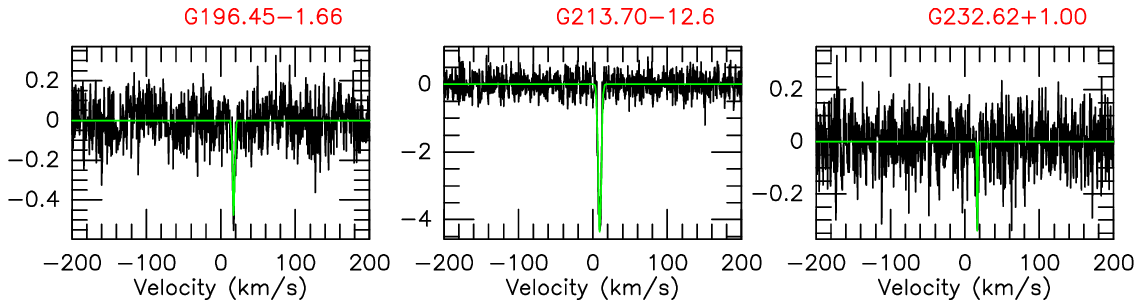}
\caption{--Continued.}
\end{figure}

\begin{figure}[h]
\centering
\includegraphics[width=17cm, height=17cm]{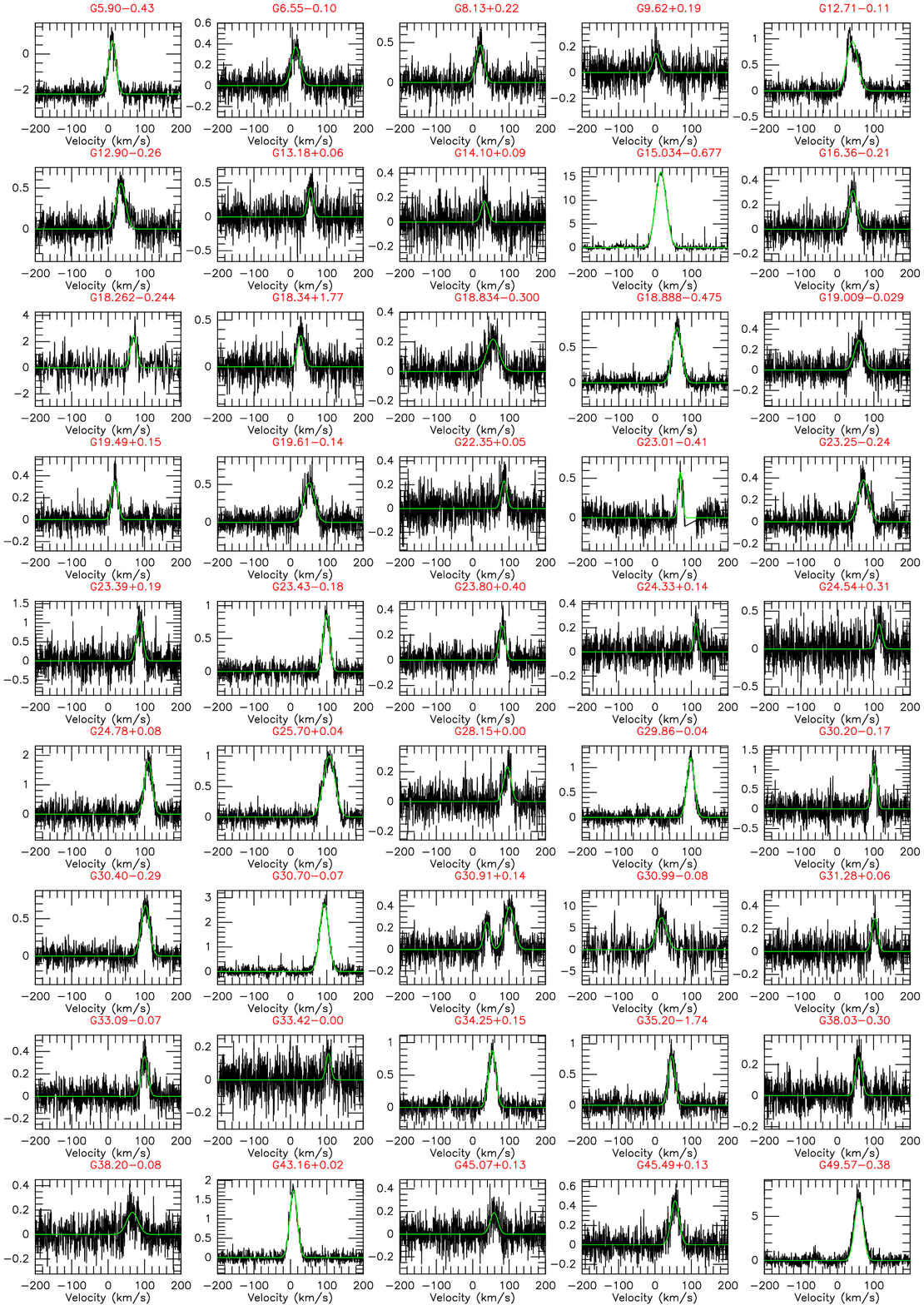}
\includegraphics[width=17cm, height=3.7cm]{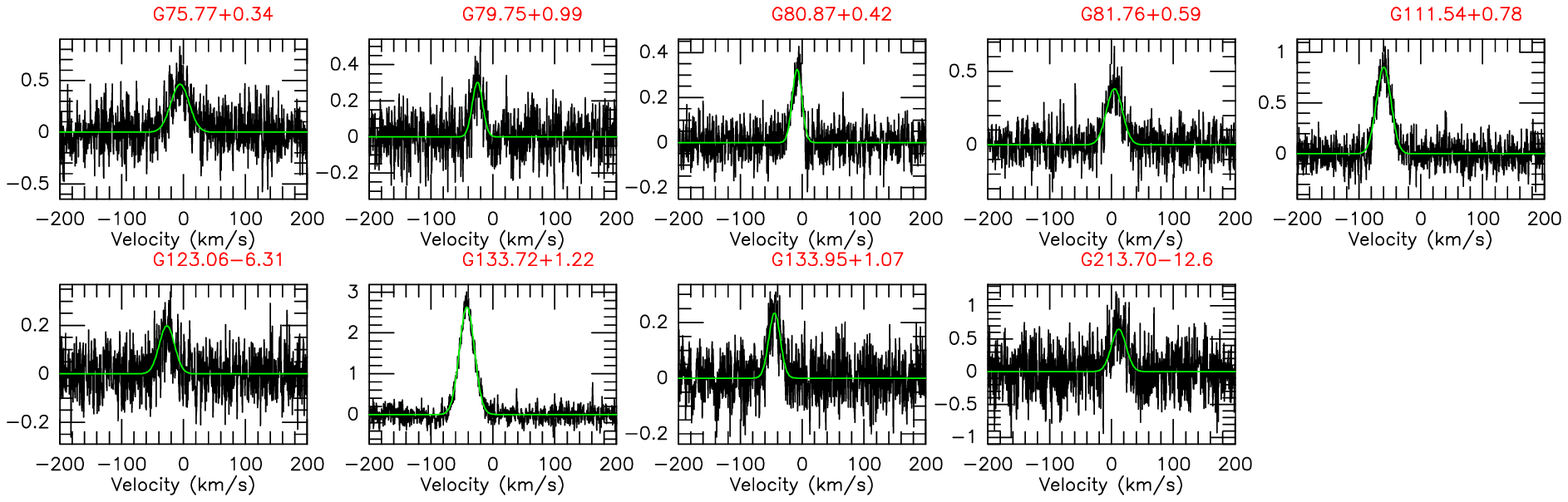}
\caption{Observed H110$\alpha$ Radio Recombination Spectral Lines (The vertical axes are Line Intensities (in Jy)).}
\end{figure}

\onecolumn
\begin{deluxetable}{lccccccc}
\tabletypesize{\scriptsize}
\tablecaption{Table of the 176 Methanol Sources observed\label{tbl-2}}
\tablewidth{0pt}
\tablehead{
Source & RA (2000) & DEC (2000) & Velocity & Intensity & Reference &
\multicolumn{1}{c}{Flag\tablenotemark{x}} & \multicolumn{1}{c}{Flag\tablenotemark{y}} \\
 & $(^{h m s})$ & ($^{\circ}$ $'$ $''$) & (km s$^{-1}$) &  (Jy) &  & for H$_2$CO & for H110$\alpha$\\}
\startdata

G121.24-0.34   &  00 36 47.5   &  +63 29 02.0   &  -22.4   &   14    &  a  &  1  &  0 \\
G123.06-6.31   &  00 52 25.1   &  +56 33 54.3   &  -29.1   &   24    &  a  &  1  &  1 \\
G133.72+1.22   &  02 25 41.8   &  +62 06 05.0   &  -44.3   &   5     &  a  &  1  &  1 \\
G133.95+1.07   &  02 27 05.0   &  +61 52 25.5   &  -44.6   &   3741  &  a  &  1  &  1 \\
G136.84+1.14   &  02 49 23.2   &  +60 47 01.2   &  -45.3   &   24    &  a  &  2  &  0 \\
G174.19-0.09   &  05 30 42.0   &  +33 47 14.0   &  2.1     &   94    &  b  &  2  &  0 \\
G173.48+2.43   &  05 39 10.3   &  +35 45 19.2   &  -13.6   &   256   &  a  &  1  &  0 \\
G173.71+2.35   &  05 39 27.6   &  +35 30 58.0   &  -14     &   5     &  b  &  2  &  0 \\
\enddata
\tablenotetext{x}{A flag of 1 means H$_2$CO detected, 0 means H$_2$CO not detected, and 2 means H$_2$CO newly detected.}
\tablenotetext{y}{A flag of 1 means H110$\alpha$ detected, 0 means H110$\alpha$ not detected}
\tablecomments{Columns 4 and 5 respectively illustrate the methanol line velocities and line intensities as obtained from references indicated on the 6th Column. The references are denoted as follows: a=Xu et al. (2003), b=Pestalozzi et al. (2005), c=Xu et al. (2009), and d=Green et al. (2010). Columns 7 and 8 are respectively flags that indicate whether or not H$_2$CO and H110$\alpha$ were detected from the sources. }
\end{deluxetable}

\begin{deluxetable}{lccc}
\tabletypesize{\scriptsize}
\tablecaption{Literature sources for previously observed 4.8 GHz formaldehyde absorptions\label{tbl-2}}
\tablewidth{0pt}
\tablehead{
S/No. & Source & S/No. & Source}
\startdata

1&	Whiteoak \& Gardner (1970)&	25&	Caswell \& Haynes (1987)\\
2&	Kutner \& Thaddeus (1971)&	26&	Heithausen et al. (1987)\\
3&	Wilson (1972)&	27&	Duncan et al. (1987)\\
4&	Scoville \& Solomon (1972)&	28&	Pipenbrink \& Wendker (1988)\\
5&	Minn \& Greenberg (1973)&	29&	Sandqvist et al. (1988)\\
6&	Gardner et al. (1973)&	30&	Wadiak et al. (1988)\\
7&	Dieter (1973)&	31&	Kogul et al. (1989)\\
8&	Whiteoak \& Gardner (1974)&	32&	Boyce et al. (1989)\\
9&	Lucas et al. (1976)&	33&	Zylka et al. (1992)\\
10&	Sandqvist \& Lindroos (1976)&	34&	Mehringer et al. (1995)\\
11&	Rieu \& Pankonin (1977)&	35&	Reynoso \& Goss (2002)\\
12&	Rickard et al. (1977)&	36&	Araya et al. (2002)\\
13&	Dickel et al. (1977)&	37&	Young et al. (2004)\\
14&	Minn \& Greenberg (1979)&	38&	Araya et al. (2004)\\
15&	Javanaud (1979)&	39&	Rodriguez et al. (2006)\\
16&	Downes et al. (1980)&	40&	Hoffman et al. (2007)\\
17&	Goss et al. (1980)&	41&	Araya et al. (2007)\\
18&	Loren (1981)&	42&	Du et al. (2011)\\
19&	Vanden-Bout et al. (1983)&	43&	Ginsburg et al. (2011)\\
20&	Wendker et al. (1983)&	44&	Wu et al. (2011)\\
21&	Poeppel et al. (1983)&	45&	Han at al. (2011)\\
22&	Goss et al. (1984)&	46&	Zhang et al. (2012)\\
23&	Gardner \& Whiteoak (1984)&	47&	Tang et al. (2013)\\
24&	Martin-Pintado et al. (1985)

\enddata
\end{deluxetable}

\begin{deluxetable}{lccccccc}
\tabletypesize{\scriptsize}
\tablecaption{Formaldehyde Line Parameters.\label{tbl-2}}
\tablewidth{0pt}
\tablehead{
Source & Velocity      & Width         & Intensity  & Flux \\
       & (km s$^{-1}$) & (km s$^{-1}$) & (Jy)       & (Jy km s$^{-1}$)}
\startdata

G2.14+0.01   &   -1.23 (0.36)   & 3.68 (0.75)   & -0.35 (0.12)   & -1.37 (0.27)   \\
G2.14+0.01   &   8.48 (0.52)   & 8.74 (1.76)   & -0.44 (0.12)   & -4.13 (0.61)   \\
G2.14+0.01   &   15.64 (0.07)   & 2.37 (0.22)   & -1.49 (0.12)   & -3.76 (0.39)   \\
G2.14+0.01   &   128.90 (0.38)   & 15.91 (0.77)   & -0.68 (0.12)   & -11.48 (0.50)   \\
G2.53+0.19   &   7.59 (0.32)   & 14.18 (0.82)   & -0.97 (0.17)   & -14.64 (0.68)   \\
G5.90-0.43   &   9.98 (0.11)   & 3.82 (0.29)   & -1.35 (0.14)   & -5.48 (0.32)   \\
G5.90-0.43   &   18.06 (0.19)   & 5.81 (0.48)   & -0.90 (0.14)   & -5.56 (0.37)   \\
G06.189-0.358   &   -32.54 (0.17)   & 2.99 (0.58)   & -0.53 (0.11)   & -1.70 (0.23)   \\
\enddata
\end{deluxetable}

\begin{deluxetable}{lccccccc}
\tabletypesize{\scriptsize}
\tablecaption{H110$\alpha$ Line Parameters.\label{tbl-2}}
\tablewidth{0pt}
\tablehead{
Source & Velocity      & Width         & Intensity  & Flux \\
       & (km s$^{-1}$) & (km s$^{-1}$) & (Jy)       & (Jy km s$^{-1}$)}
\startdata

G5.90-0.43   &   10.99 (0.37)   & 24.19 (0.90)   & 1.18 (0.19)   & 30.38 (0.96)   \\
G6.55-0.10   &   14.39 (0.65)   & 28.17 (1.53)   & 0.37 (0.09)   & 11.09 (0.52)   \\
G8.13+0.22   &   19.95 (0.63)   & 24.82 (1.55)   & 0.47 (0.12)   & 12.32 (0.64)   \\
G9.62+0.19   &   3.11 (1.29)   & 21.86 (3.68)   & 0.15 (0.09)   & 3.49 (0.45)   \\
G12.71-0.11   &   41.08 (0.39)   & 36.37 (0.86)   & 0.95 (0.13)   & 36.72 (0.78)   \\
G12.90-0.26   &   34.71 (0.58)   & 30.20 (1.57)   & 0.56 (0.12)   & 17.93 (0.72)   \\
G13.18+0.06   &   54.92 (0.80)   & 17.85 (1.88)   & 0.43 (0.17)   & 8.17 (0.75)   \\
G14.10+0.09   &   32.51 (1.44)   & 21.74 (3.69)   & 0.17 (0.11)   & 3.89 (0.54)   \\
\enddata
\end{deluxetable}

\begin{deluxetable}{cccccccc}
\tabletypesize{\scriptsize}
\tablecaption{Detection rates of the H$_2$CO absorption lines.
\label{tbl-2}}
\tablewidth{0pt}
\tablehead{
CH$_3$OH Intensity & Number of CH$_3$OH & Number of H$_2$CO  & H$_2$CO \\
bands & sources observed &  sources detected & detection rates  }
\startdata

I $\geq$ 100 Jy           & 30 & 26 & 86.7\% \\
50 Jy $\leq$ I $<$ 100 Jy & 23 & 21 & 91.3\% \\
10 Jy $\leq$ I $<$ 50 Jy  & 88 & 67 & 76.1\% \\
5 Jy $\leq$ I $<$ 10 Jy   & 35 & 24 & 68.6\% \\

\enddata
\end{deluxetable}

\begin{deluxetable}{lccccccc}
\tabletypesize{\scriptsize}
\tablecaption{H$_2$CO Association with H110$\alpha$, EGO, IRDC and IRAS sources. The number 1 denotes an association while the number 0 denotes a non-association.
\label{tbl-2}}
\tablewidth{0pt}
\tablehead{
H$_2$CO sources & Association with H110$\alpha$ & Association with EGOs & Association with IRDCs & Association with IRAS sources  }
\startdata

G2.14+0.01       &      0	 &  0	 &  0	 &  1   \\
G2.53+0.19       &      1	 &  0	 &  0	 &  0   \\
G5.90-0.43       &      1	 &  0	 &  0	 &  1   \\
G06.189-0.358    &      0	 &  0	 &  0	 &  0   \\
G6.55-0.10       &      1	 &  0	 &  0	 &  1   \\
G6.78-0.27       &      1	 &  0	 &  0	 &  1   \\
G07.601-0.139    &      0	 &  0	 &  0	 &  1   \\
G8.13+0.22       &      1	 &  0	 &  0	 &  0   \\
\enddata
\end{deluxetable}

\end{document}